\documentclass[11pt,a4paper]{iopart}
\usepackage{iopams}
\usepackage[english]{babel}
\usepackage{float}
\usepackage[dvips]{graphicx}
\usepackage{url}
\usepackage{rotating}
\newcommand{\gadget}{{\footnotesize GADGET}}
\usepackage{epsfig}
\usepackage[dvips]{color}
\usepackage{listings}

\definecolor{Red}{named}{Red}

\begin{document}
\title{Neutrinos in Non-linear Structure Formation - a Simple SPH Approach}
\author{Steen Hannestad$^1$, Troels Haugb{\o}lle$^2$, Christian Schultz$^1$}
\ead{sth@phys.au.dk, haugboel@nbi.dk, cs06@phys.au.dk}
\address{$^1$~Department of Physics and Astronomy, Aarhus University, Ny Munkegade, DK-8000 Aarhus C, Denmark. \\
$^2$~Centre for Star and Planet Formation, Natural History Museum of Denmark, University of Copenhagen, {\O}ster Voldgade 5-7, DK-1350 Copenhagen, Denmark.}
\date{\today}

%%%%%%%%%%%%%%%%%%%%%%%%%%%%%%%%%%%%%%%%%%%%%%%%%%%%%%%%%%%%%%%%%%%%%%%%%%%%%%%%%%%%%%%%%%%%%%%%%%%%%%%%%%%%%%%%%%%%%%%%%%%%%%%%%%%%%
\begin{abstract}
We present a novel method for implementing massive neutrinos in $N$-body simulations. Instead of sampling the neutrino velocity distribution by individual point particles we take neutrino free-streaming into account by treating it as an effective redshift dependent sound speed in a perfect isothermal fluid, and assume a relation between the sound speed and velocity dispersion of the neutrinos. Although the method fails to accurately model the true neutrino power spectrum, it is able to calculate the total matter power spectrum to the same accuracy as more complex hybrid neutrino methods, except on very small scales. We also present an easy way to update the publicly available \gadget-2 version with this neutrino approximation.
\end{abstract}

\section{Introduction}

Over the past few years cosmological measurements have become much more precise and the need for correspondingly accurate theoretical predictions is rapidly growing. Upcoming measurements of weak gravitational lensing and galaxy clustering at high redshift are expected to reach percent level accuracy on scales where non-linear gravitational effects are important. On these scales there are a number of difficult challenges that must be overcome before percent level precision in the theoretical models can reliably be obtained. For example, baryonic effects must be properly taken into account (see e.g.\ \cite{Heitmann:2008eq} for a recent discussion of the required accuracy).
Another important issue is the presence of massive neutrinos in the universe. Even though neutrinos are very light, they do contribute to cosmic structure formation and have an impact which is certainly larger than the accuracy goal.
This has led to a number of analytic or semi-analytic studies of non-linear gravitational clustering with neutrinos, using either higher order perturbation theory or renormalisation group methods inspired by field theory
\cite{Lesgourgues:2009am,Saito:2009ah,Wong:2008ws,Saito:2008bp}. Even though such methods are extremely important, in particular in the pseudo linear regime, they must be checked and/or calibrated using high precision $N$-body simulations, and they are hard to combine self consistently with $N$-body models including baryon physics.

Early simulations including massive neutrinos were exclusively done in mixed dark matter models with high neutrino masses and no dark energy (see e.g.\ \cite{Kofman:1995ds}). However, observations strongly suggest that the underlying cosmological model is a variant of the $\Lambda$CDM model with relatively modest neutrino masses of approximately 1 eV or less.
Of course it should be noted that current data actually favours more than the 3 neutrino degrees of freedom predicted by the standard model (see e.g.\ \cite{Komatsu:2010fb,Hamann:2011ge,Hamann:2010bk,Giusarma:2011ex,Hou:2011ec,Keisler:2011aw,GonzalezMorales:2011ty,Kristiansen:2011mp}), and that an additional sterile neutrino with a mass around 1 eV might explain this, as well as the observed short baseline neutrino oscillation anomalies (see e.g.\ \cite{Giunti:2011gz,Giunti:2011hn,Kopp:2011qd} for recent discussions of the short baseline anomaly).

Neutrinos are well known to be difficult to handle in structure formation simulations. The reason is that for reasonable values of the neutrino mass, the thermal velocities of typical neutrinos vastly exceed the gravitational flow velocities until very late in the evolution of the universe. This in turn leads to very short time steps and thermal poisson noise in the simulations unless very large numbers of neutrino particles are used.
This problem has been addressed in a number of different ways, for example by evolving the neutrino density on a grid, or by a combination of grid and particle methods. While such simulations can achieve high accuracy they are cumbersome and complex to run. We refer the reader to a number of papers discussing these issues
\cite{Brandbyge1,Brandbyge2,Brandbyge3,Brandbyge4,Viel:2010bn,Viel2011,Agarwal:2010mt}.

Here we present a much simpler approach, based on the assumption that neutrinos can be treated as a perfect fluid with a finite sound speed. While this assumption is manifestly wrong, we demonstrate that quantities such as the matter power spectrum can be calculated with high accuracy (reaching an accuracy of 1-2\% over the relevant range of scales) in such simulations which have the advantage of being very fast and simple to run.

In the next section we briefly outline the theoretical framework for treating neutrinos in structure formation. In Section 3 we describe our implementation on massive neutrinos as a fluid in the publicly available \gadget-2 code, and in Section 4 we provide a detailed description of our results, including convergence tests. Finally Section 5 contains a discussion and our conclusions. \ref{nusph} describes the small changes necessary to add neutrino particles to the latest public version of the \gadget-2 code. A link to a working version is also provided.

\section{Theory}\label{theory}
\subsection{The Boltzmann equation}
In this subsection we will briefly outline how the evolution of massive neutrinos is followed in linear theory. The notation is identical to \cite{Ma}. We use the metric in the conformal Newtonian gauge

\begin{equation}
  ds^2 = -a^2(1+2\psi)d\tau^2 + a^2(1-2\phi)dx^2.
  \label{eq:metric}
\end{equation}

In a perturbed universe the phase-space distribution function is expanded to first order as follows
\begin{equation}
  f= f_0 + \frac{\partial f_0}{\partial T} \delta T = f_0(1+\Psi),
  \label{eq:phase_space}
\end{equation}
with the perturbation parametrised by $\Psi = - d {\rm ln} f_0 / d {\rm ln} q ~ \delta T / T$, and the zeroth-order Fermi-Dirac phase-space distribution function given by
\begin{equation}
  f_0(q) = \frac{1}{e^{q/T} + 1}.
  \label{eq:f0}
\end{equation}
Here $q_i = a p_i$, where $p_i$ is the proper redshifting momentum and $T$ is the neutrino temperature today.

After neutrino decoupling the collisionless Boltzmann equation evolves the distribution function as

\begin{equation}
  \frac{df}{d\tau} = \frac{\partial f}{\partial \tau} + \frac{dx^i}{d\tau}\frac{\partial f}{\partial x^i}
   + \frac{dq}{d\tau}\frac{\partial f}{\partial q} + \frac{dn_i}{d\tau}\frac{\partial f}{\partial n_i} = 0.
  \label{eq:Boltzmann}
\end{equation}

Expanding the perturbation $\Psi$ in a Legendre series the perturbed neutrino energy density is given by a weighted sum over neutrino momentum states

\begin{equation}
  \label{eq:drho}
  \delta \rho_\nu (k) = 4\pi a^{-4} \int{q^2dq\epsilon f_0 \Psi_0},
\end{equation}
where $\epsilon = (q^2 + a^2 m^2)^{1/2}$.

From the Boltzmann equation the $\Psi_l$'s are related to each other and the metric potentials by
\begin{eqnarray}
\label{eq:Psi_l0}
\dot\Psi_0 & = & -\frac{qk}{\epsilon}\Psi_1 - \dot\phi \frac{d{\rm ln}f_0}{d{\rm ln}q}, \\
\dot\Psi_1 & = & \frac{qk}{3\epsilon} \biggl(\Psi_0 - 2\Psi_2 \biggr) - \frac{\epsilon k}{3q}\psi \frac{d{\rm ln}f_0}{d{\rm ln}q},\\
\dot\Psi_l & = & \frac{qk}{\epsilon(2l+1)} \biggl(l\Psi_{l-1} - (l+1)\Psi_{l+1} \biggr),~l\ge 2.
\label{eq:Psi_l2}
\end{eqnarray}
The second term on the right hand side of the equation for $\dot\Psi_0$ encodes the effect of structure formation in evolving gravitational potentials, whereas the first term incorporates the effect of velocity on structure formation. The change in velocity is affected by the three terms on the right hand side of the equation for $\dot\Psi_1$. The first term encodes the effect of flow velocities redshifting in an expanding Universe, whereas the second term, found from the hierarchy of $\dot\Psi_l$'s, incorporates the effect of momentum (and redshift) dependent neutrino free-streaming. These two terms give rise to less structure, whereas the last term in the equation for $\dot\Psi_1$ gives the acceleration as a gradient of the gravitational potential.

The above set of hierarchy equations can formally be integrated to give the fluid equations \cite{Ma}
\begin{eqnarray}
\label{eq:fluid}
\dot \delta & = & -(1+w)(\theta - 3 \dot \phi)-3 \frac{\dot a}{a} \left(\frac{\delta P}{\delta \rho}-w\right) \delta \\
\dot \theta & = & - \frac{\dot a}{a}(1-3w) \theta -\frac{\dot w}{1+w} \theta + \frac{{\delta P}/{\delta \rho}}{1+w} k^2 \delta - k^2 \sigma + k^2 \psi,
\end{eqnarray}
where $\delta$ is the density contrast, $\theta$ is the fluid velocity perturbation, $w=P/\rho$ is the equation of state, and $\sigma$ is the anisotropic stress.

For a perfect fluid we have $\sigma = 0$ and the hierarchy can be truncated at $l=1$. This reduces the Boltzmann equation to the usual continuity and Euler equations. Even though this is formally not allowed for massive neutrinos the system of equations could still be truncated by assuming a relation between $\sigma$ and the effective sound speed, $c_s$, in the system. This approach was taken in \cite{shoji} and shown to provide reasonable accuracy when tracking the linear evolution of neutrino perturbations.
Since neutrinos are a subdominant component in structure formation (except in sterile neutrino warm dark matter scenarios), a relatively small error in the neutrino component translates into a much smaller error in such quantities as the total matter power spectrum.

In the neutrino fluid approach it is assumed that neutrinos behave like a perfect fluid with sound speed squared, $c_s^2$, given by
\begin{equation}\label{eq:cs}
c_s^2 = \alpha F(z) \sigma_\nu^2,
\end{equation}
where $\sigma_\nu$ is the velocity dispersion of neutrinos, given by their effective temperature, and $F(z)$ is some function of redshift (but not of position). $\alpha$ is a constant of order 1.
In the non-relativistic limit we have $c_s^2 \ll 1$, $w \sim 0$, and the above equations reduce to
\begin{eqnarray}
\dot \delta & = & -(\theta - 3 \dot \phi) \\
\dot \theta & = & -\frac{\dot a}{a} \theta + c_s^2 k^2 \delta + k^2 \psi.
\end{eqnarray}

In the non-relativistic limit the neutrino velocity dispersion can be written as
\begin{equation}
\sigma_\nu^2 = \frac{\int p^2/m_\nu^2f_0(p)d^3p}{\int f_0(p)d^3p}
\end{equation}
which in the same limit reduces to the following simple form \cite{shoji}
\begin{equation}\label{eq:cs2}
\sigma_\nu^2 = \frac{15 \zeta(5)}{\zeta(3)} \left(\frac{4}{11}\right)^{2/3} \frac{T_\gamma^2 (1+z)^2}{m_\nu^2},
\end{equation}
and $F(z) \simeq (5/9)$.

Again, we stress that this approach is manifestly not correct in the sense that a perfect fluid with finite sound speed exhibits acoustic oscillations on scales smaller than the Jeans scale because of the $c_s^2 k^2 \delta$ term in the equation for $\dot \theta$ while the true neutrino equations have a $k^2 \sigma$ damping term which is not proportional to the local density. Since the Jeans scale for collisional particles is given as
\begin{equation}
k_J=\sqrt{\frac{3}{2}}\frac{H}{c_s (1+z)}
\end{equation}
and the free-streaming scale for neutrinos is given as
\begin{equation}
k_{FS}=\sqrt{\frac{3}{2}}\frac{H}{\sigma_\nu (1+z)}
\end{equation}
the relation in Eq.~(\ref{eq:cs}) ensures that the Jeans scale is close to the free-streaming scale provided that $\alpha F(z) \sim \cal{O}$(1).

However, since the neutrino transfer function at the starting point of the $N$-body simulation has been evolved using the exact Boltzmann equation hierarchy up to a point where the neutrino transfer function on small scales is orders of magnitude below the CDM transfer function, the most important point is {\it not} the difference between dissipation and oscillations, but just the fact that the neutrino transfer function cannot grow on small scales.

\section{Implementation in $N$-body simulations - $\nu$SPH}

\begin{figure}
     \noindent
     \vspace*{-1cm}
     \begin{center}
           \includegraphics[width=0.7\linewidth]{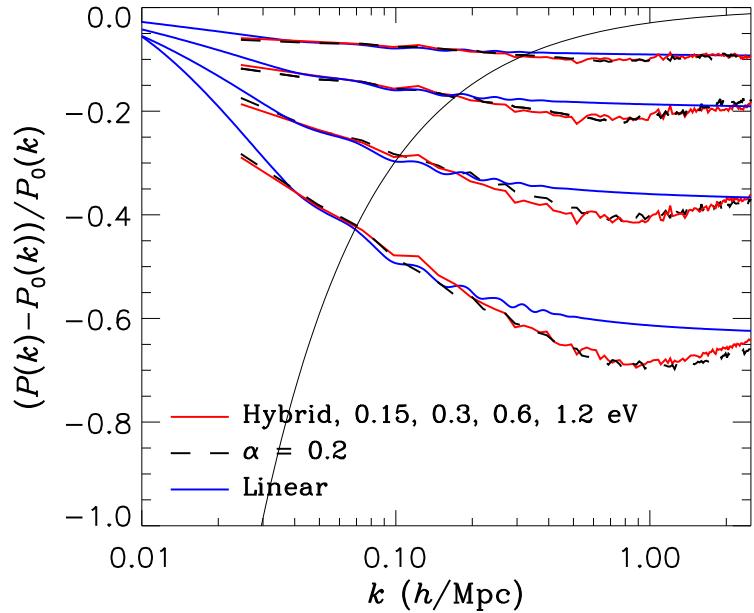}
     \end{center}
     \caption{Power spectrum damping for various neutrino masses. Thin black line is the estimated $1\sigma$ uncertainty on $P(k)$ from a galaxy survey with effective volume $V_{\rm eff} =$ 10 (Gpc/$h)^3$ \cite{tegmark}.}
   \label{fig:power}
\end{figure}

\subsection{Parameters and initial conditions}

Our simulations use the same cosmological parameters as in our previous studies, i.e.\ they assume a flat, $\Lambda$CDM model with $\Omega_b = 0.05$, $\Omega_m = 0.3$, $\Omega_\Lambda = 0.7$ and $h = 70$. The scalar fluctuation amplitude is assumed to be $A_s = 2.3 \times 10^{-9}$. While these values are not preferred by current cosmological data, they are sufficiently close to the best fit values to produce qualitatively identical results. We have used a modified version of the \gadget-2 Particle-Mesh N-body code \cite{Springel:2000yr,Springel:2005mi} to perform the simulations.
Unless otherwise stated, our simulations are performed in boxes of size 256 Mpc/$h$, using $512^3$ particles and a $512^3$ force mesh. This is more than adequate for our purposes and we have carefully checked that resolution is not a problem in our simulations.

Initial conditions are generated using CAMB to evolve the linear transfer functions until a redshift of $z=49$, at which point a 2nd order Lagrangian method is used to generate the perturbation field, as described in \cite{Brandbyge1}. We assume adiabatic fluctuations.

\subsection{Neutrinos}

We implement neutrinos in \gadget-2 as a new SPH particle type. However, instead of the normal SPH equations we assume an isothermal neutrino gas where the only free parameter is the redshift dependent sound speed. This parameter is simply found by using the relation in Eq.~(\ref{eq:cs}). Given that in the normal gas formulation the isothermal sound speed is redshift independent this has to be encoded in the equations of motion for the SPH particles. In \gadget-2 a constant entropy formulation is used that allows for change in entropy only at shocks or due to external physics. We could use the explicit equations (\ref{eq:cs}) and (\ref{eq:cs2}), but for simplicity we instead use them to set the initial entropy, and then evolve the entropy forward using the simple evolution equation
\begin{equation}
\frac{d \textrm{A}}{d\textrm{T}} = -2\, \textrm{A} / \textrm{T}\,,
\end{equation}
where $\textrm{T}=a$ is the time variable and $\textrm{A}=c_s^2$ is the effective entropy in \gadget-2. The treatment of neutrinos as a fluid induces acoustic oscillations at small scales. We do not want the oscillations to be reflected in the gravitational potential, and therefore it is essential to set the gravitational smoothing large enough that they do not impact the CDM. This is discussed below. Finally we note that although we have used SPH as a framework for treating neutrinos in the non-linear regime, a mesh implementation should work equally well.

\section{Results}

\subsection{Power spectra}

\begin{figure}
     \noindent
     \vspace*{0cm}
     \begin{center}
           \includegraphics[width=1.0\linewidth]{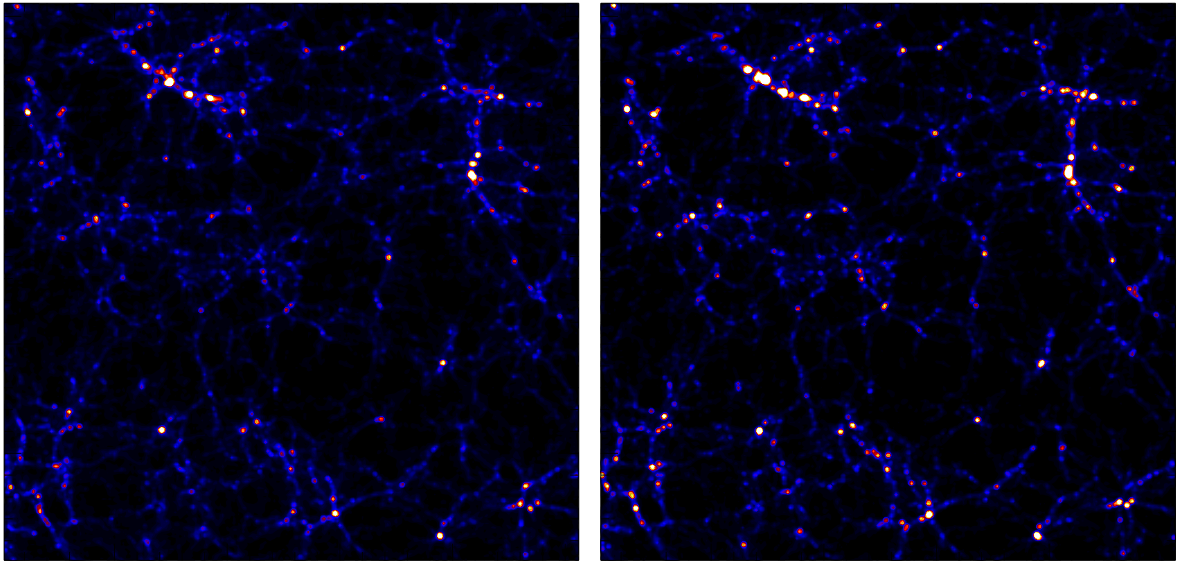}
     \end{center}
     \caption{Density in the CDM component for a neutrino mass of 1.2 eV (left) and 0.15 eV (right). Simulations are made with $512^3$ particles in a 256 Mpc/$h$ box, and the figures show slices of thickness 10 Mpc/$h$.}
   \label{fig:dens}
\end{figure}

The total matter power spectra can reproduce results from the detailed hybrid code \cite{Brandbyge3} to a precision of approximately 1-2\% up to $k \sim 1-2$, more than sufficient for analysing most upcoming data sets.
In Fig.~\ref{fig:power} we show the suppression of fluctuation power as a function of different neutrino masses up to 1.2 eV. The results are compared to the exact results from the hybrid code, and to results from linear perturbation theory. To provide an estimate of the maximum deviation tolerable we also show the statistical error on a power spectrum measurement from a galaxy survey with an effective volume of 10 (Gpc/$h)^3$. From the figure it can be seen that deviations between the $\nu$SPH code and the hybrid codes are smaller than the statistical uncertainty at least up to $k \sim 1-2$.
For illustration purposes we show CDM density plots from simulations with $m_\nu = 0.15$ eV and 1.2 eV respectively in Fig.~\ref{fig:dens}.

However, the neutrino clustering is not reproduced to the same level of accuracy. As is the case for grid-based simulations, an important ingredient has been removed.
Decoupled neutrinos have a distribution given by a relativistic Fermi-Dirac function. This means that even though the mean momentum is large, a small proportion of neutrinos will always have low initial momenta and be subject to gravitational clustering. This phenomenon was studied in great detail in \cite{Brandbyge4} and the neutrino density profiles in typical dark matter halos extracted.
However, in the $\nu$SPH approach the neutrino gas is treated as isothermal which in particle language is approximately equal to providing all neutrinos with the same initial thermal velocity. While this is a reasonable approximation to the real physical situation it fails to follow the small proportion of low momentum neutrinos that actually cluster in halos.

In terms of the neutrino power spectrum this can be seen as a lack of power at high $k$. Again, we stress that this is not a problem in most cases, i.e.\ where the total matter power spectrum is studied: weak lensing, large scale structure surveys, Lyman-$\alpha$ surveys etc. However, the code is not well suited for studies of neutrino behaviour, for example in the context of relic neutrino detection. In practise the capability to follow the detailed neutrino distribution is rarely needed and the ease with which the $\nu$SPH method can be implemented far outweighs its limitations.

\begin{figure}
     \noindent
     \vspace*{-1cm}
     \begin{center}
           \includegraphics[width=0.7\linewidth]{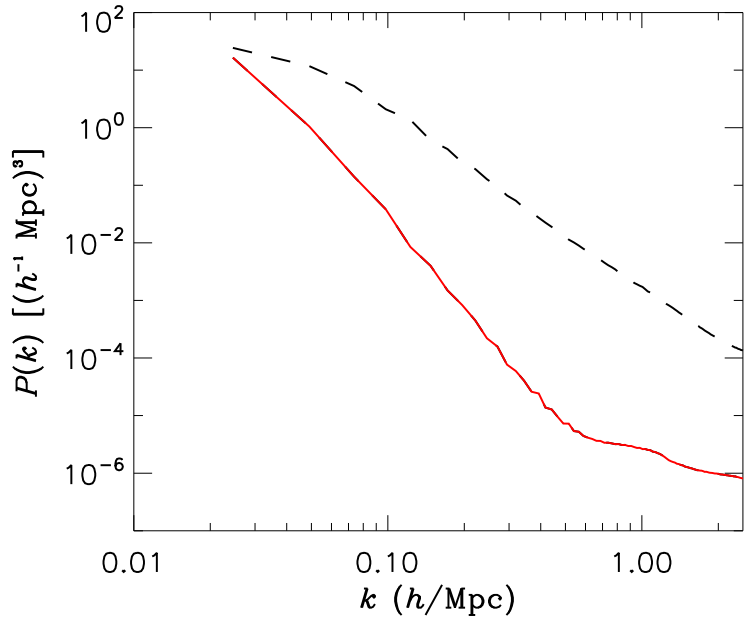}
     \end{center}
     \caption{Neutrino power spectra at $z=0$ for 0.6 eV neutrinos in the $\nu$SPH approximation (black line) and in the hybrid code (red line).}
   \label{fig:nupower}
\end{figure}

\subsubsection{Evolution with redshift}

\begin{figure}
     \noindent
     \vspace*{-1cm}
     \begin{center}
           \includegraphics[width=0.7\linewidth]{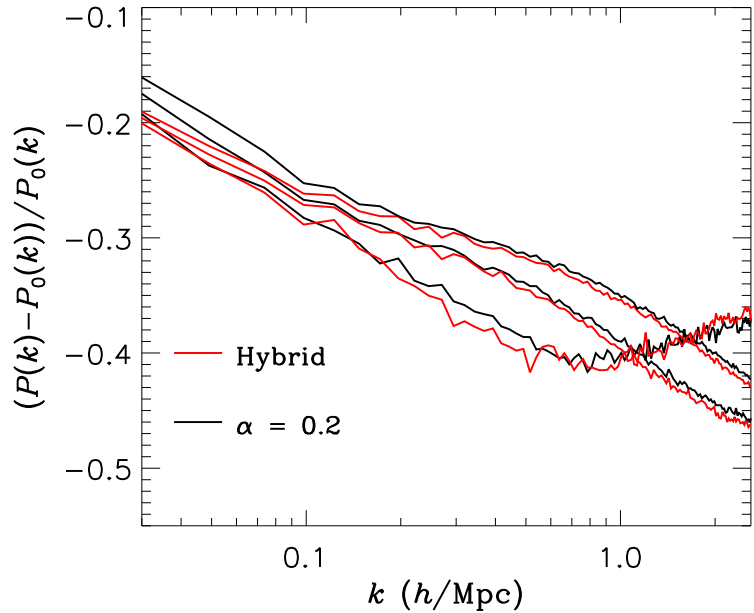}
     \end{center}
     \caption{Power spectrum suppression at $z=4, 7/3, 0$ (in order of decreasing amplitude on large scales) for 0.6 eV neutrinos in the $\nu$SPH approximation.}
   \label{fig:06evz}
\end{figure}

The pronounced dip in suppression is located at the point where the $\Lambda$CDM model has gone non-linear while the $\nu \Lambda$CDM model is still in the linear regime. At higher redshifts this point is located at higher values of $k$, as can be seen in Fig.~\ref{fig:06evz}. The amplitude of the suppression also increases because the overall power spectrum normalisation is lower at high redshift. We have checked explicitly that the $\nu$SPH code produces results which are compatible with the hybrid code. Fig.~\ref{fig:06evz} clearly shows that the agreement between the two codes is even better at higher redshifts than at $z=0$.

\subsubsection{Gas softening}

\begin{figure}
     \noindent
     \vspace*{-1cm}
     \begin{center}
           \includegraphics[width=0.7\linewidth]{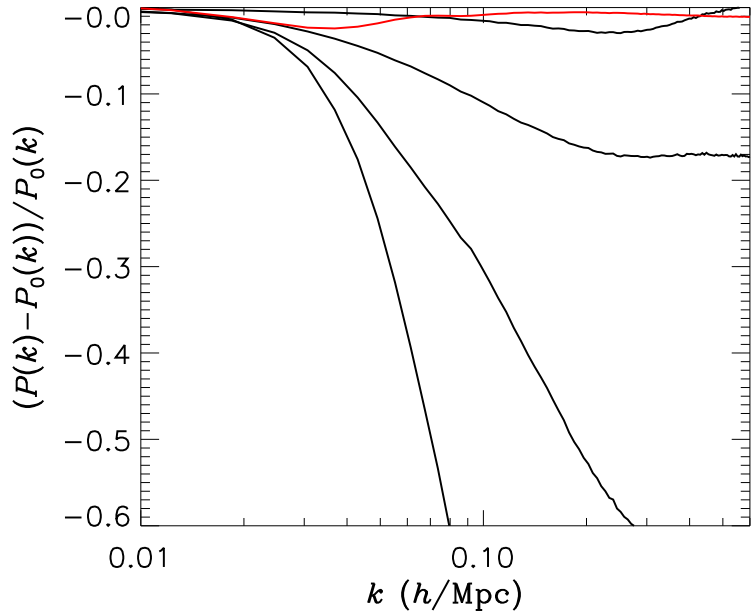}
     \end{center}
     \caption{The effect of gas softening in the $\nu$SPH simulations. Black lines are $256^3$ simulations in a $1024$ Mpc$/h$ box. The lines show the matter power spectrum for $r_{\rm soft} = 2000, 1000, 500, 250$ kpc, relative to a model with $r_{\rm soft} = 4000$ kpc. The red line shows the same for $r_{\rm soft} = 1000$ kpc relative to $r_{\rm soft} = 4000$ kpc, but for a $512^3$ simulation. As can be seen, the effect diminishes dramatically when a larger number of particles is used.}
   \label{fig:soft}
\end{figure}

\begin{figure}
     \noindent
     \vspace*{0cm}
     \begin{center}
           \includegraphics[width=1.0\linewidth]{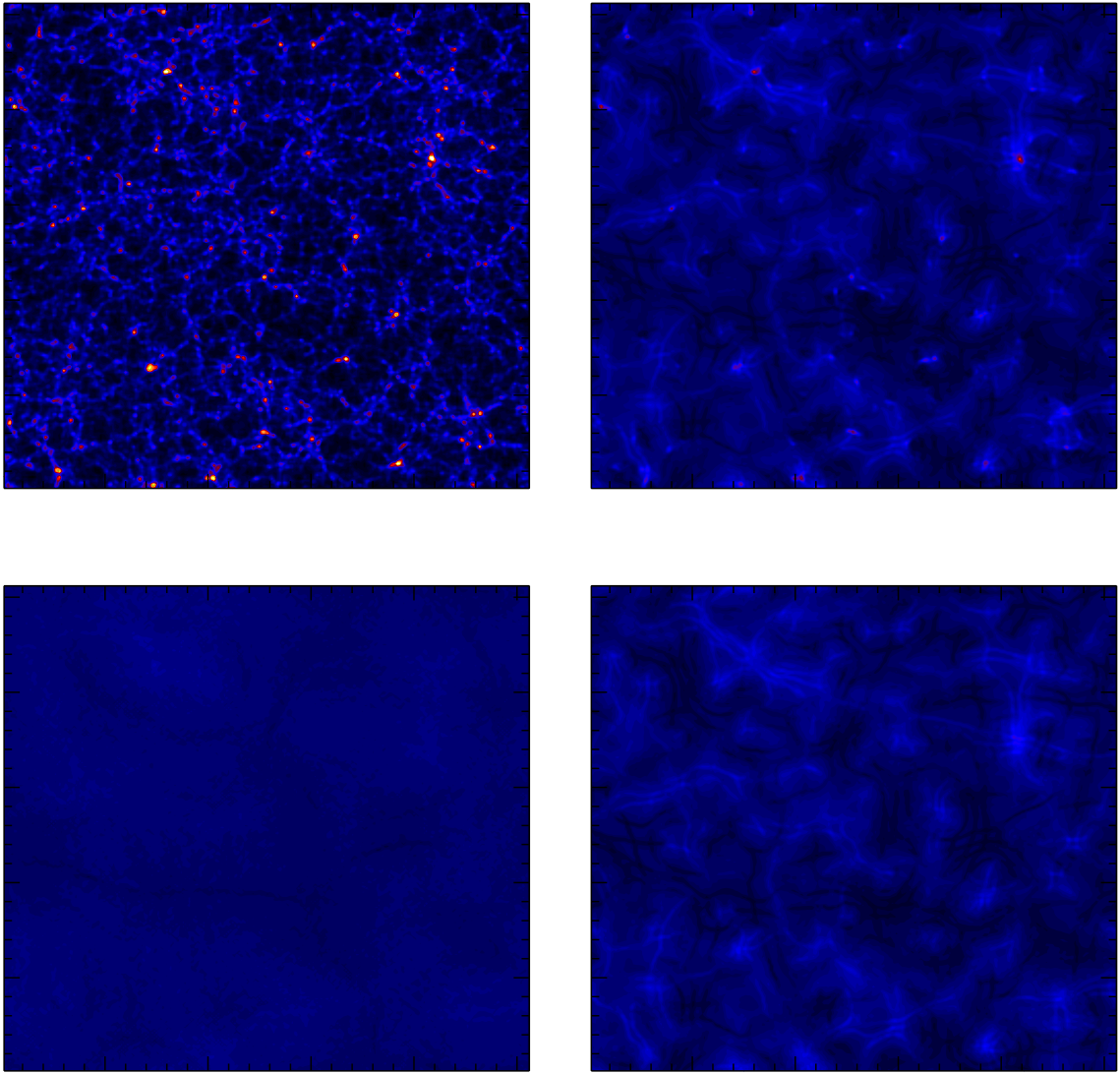}
     \end{center}
     \caption{Density in simulations with 0.6 eV neutrinos in a 1024 Mpc/$h$ box using $256^3$ resolution. The plot shows a slice of thickness 40 Mpc/$h$. Top row is CDM, bottom neutrinos. The left column is with an SPH softening of 4000 kpc/$h$, and the right with a softening of 250 kpc/$h$. The acoustic waves are seen in both the neutrino and the CDM components in the right column.}
   \label{fig:waves}
\end{figure}

One important effect which must be taken into account is that if the effective softening length for SPH particles is set too low, the system fails to properly track evolution not only of the SPH system, but also the dominant CDM component.
This can be seen in the total matter power spectrum in Fig.~\ref{fig:soft}. The matter power spectrum drops steeply on scales smaller than the effective sound propagation scale. That this is indeed due to waves can be seen in Fig.~\ref{fig:waves} where the right side panels clearly show oscillations in both the CDM and the neutrino components. What happens is that the strong waves in the neutrino component wipe out the CDM fluctuations before they become highly non-linear, except in a few very high density regions.

When the softening length is increased the spurious acoustic waves vanish and the CDM system behaves as expected. We stress that in the simulations actually used to compare against the hybrid code, this is not an issue. We merely show Figs.~\ref{fig:soft} and \ref{fig:waves} for illustration purposes.

\subsubsection{Effective sound speed}

\begin{figure}
     \noindent
     \vspace*{-1cm}
     \begin{center}
           \includegraphics[width=0.7\linewidth]{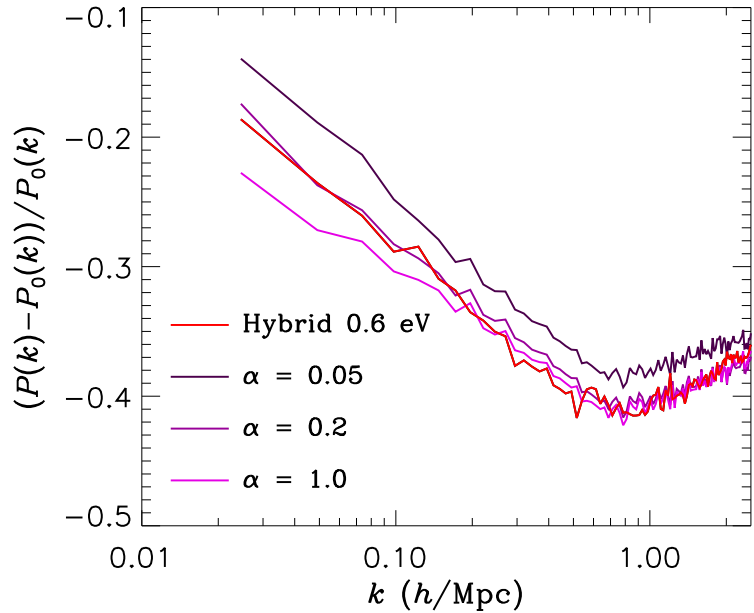}
     \end{center}
     \caption{The relative suppression in a model with 0.6 eV neutrinos.}
   \label{fig:velo}
\end{figure}

In Fig.~\ref{fig:velo} we show the relative suppression for 0.6 eV for different values of the effective sound speed. For low values the overall suppression is systematically underestimated. However, for values around 0.2 and up, the small scale structure fits extremely well. There is a simple reason for this: On small scales neutrinos contribute almost nothing to the overall structure. Even if they do cluster in halos, the fraction of halo masses in neutrinos is tiny. Therefore it is a good approximation when calculating global quantities to treat neutrinos as non-clustering on sufficiently small scales. This behaviour is captured by our treatment and even though the method significantly under predicts the amount of power in the neutrino component on small scales it has no effect on quantities such as the power spectrum.

On larger scales, the effective velocity must be tuned to fit with the linear theory prediction. Simulations with too high a sound speed underestimate power on large scales, while the opposite is true for neutrino components too similar to CDM.

In conclusion: Quantities such as the power spectrum are relatively robust against variations in the actual value of the effective sound speed on small scales, while on large scales it should be tuned to fit the linear theory prediction. In practise we find a value of $\alpha$ around 0.2-0.3 to be accurate for any neutrino mass up to 1.2 eV.

\section{Discussions and conclusions}

We have presented a very simple scheme for implementing neutrinos in $N$-body simulations using SPH and the assumption that neutrinos behave as an ideal, isothermal fluid. While this assumption is wrong because neutrinos in reality are free-streaming, we have demonstrated that errors in quantities such as the matter power spectrum can be kept under control at the 1-2\% level over all relevant scales, sufficient for analysing for example data from the upcoming EUCLID satellite mission. The reason is that although neutrino small scale clustering is severely underestimated in the $\nu$SPH code, the contribution of neutrinos to small scale matter clustering is negligible and quantities such as the power spectrum can therefore be estimated quite accurately.

The SPH implementation presented here has a number of advantages over the hybrid neutrino code presented in \cite{Brandbyge3}: It is very easy to implement in existing codes, such as \gadget-2 (see \ref{nusph} for how to implement it). The only requirement is the addition of a new SPH particle type with a modified time-temperature relation. As for initial conditions, the only required input is a set of transfer functions for CDM, baryons and neutrinos. This is in contrast to the hybrid code which requires solving the linear theory neutrino Boltzmann equations either separately or in the $N$-body code. The hybrid code also necessitates solving the momentum dependent Boltzmann equations and keeping this information as the $N$-body code is evolved forward in time.

While the $\nu$SPH code fails to accurately map neutrino structures on small scales, it is more than adequate for power spectrum calculation. This in turn means that it is well suited for calculations of e.g.\ weak lensing. Cluster abundances can also be reliably calculated. As was demonstrated in \cite{Brandbyge4}, the halo mass function in models with neutrinos can be accurately calculated by using the correct power spectrum and making the assumption that neutrinos do not contribute significantly to halo masses, i.e.\ exactly the assumption needed for the SPH approach to work.

The only real drawback of the code is that it cannot follow neutrino structures. However, this feature is only really needed for calculations of quantities such as the local neutrino density.

In summary, the code presented here is more than sufficiently accurate and extremely easy to implement in existing $N$-body codes. For this reason the $\nu$SPH method could well become the standard method for implementing neutrinos in structure formation simulations.

%%%%%%%%%%%%%%%%%%%%%%%%%%%%%%%%%%%%%%%%%%%%%%%%%%%%%%%%%%%%%%%%%%%%%%%%%%%%%%%%%%%%%%%%%%%%%%%%%%%%%%%%%%%%%%%%%%%%%%%%%%%%%%%%%%%%%%
%%%%%%%%%%%%%%%%%%%%%%%%%%%%%%%%%%%%%%%%%%%%%%%%%%%%%%%%%%%%%%%%%%%%%%%%%%%%%%%%%%%%%%%%%%%%%%%%%%%%%%%%%%%%%%%%%%%%%%%%%%%%%%%%%%%%%%
\section*{Acknowledgements}
We acknowledge computing resources from the Danish Center for
Scientific Computing (DCSC).
\appendix
\section{Including SPH neutrinos into \gadget-2}
\label{nusph}

Patching a stock version of \gadget-2 with SPH neutrinos is very easy. It is only a matter of inserting $\simeq 100$ lines of code. A stock version of \gadget-2 patched with SPH neutrinos can be found on \url{http://phys.au.dk/~steen/nusph.tar.gz} - but note that here the neutrinos replace baryons as particle type 0. This means that in this version it is not possible to have both baryons and neutrinos, however this is easy to ameliorate should one wish to. A normal \gadget-2 makefile can be used to compile this code, the only modification is to add OPT+=-DNEUTRINO\_FLUID. Without this option the code behaves like the stock version. It is also necessary to add an OmegaNeutrino value in the parameter file.

The above minimal SPH neutrino code is based on \gadget-2 version 2.0.7, so to see the changes it is possible to do a simple
\begin{lstlisting}
diff -r /path/to/Gadget2-2.0.7_nusph/Gadget2 /path/to/Gadget2-2.0.7/Gadget2
\end{lstlisting}
between this version and a stock version of the \gadget-2 code. It is not too much work to put in the extra code by hand in a modified \gadget-2 code. It should also be possible to run a three-way merge, \textit{e.g.} diff3, on each of the files allvars.h, begrun.c, global.c, hydra.c, init.c, io.c and read\_ic.c and handle any conflicts.

%%%%%%%%%%%%%%%%%%%%%%%%%%%%%%%%%%%%%%%%%%%%%%%%%%%%%%%%%%%%%%%%%%%%%%%%%%%%%%%%%%%%%%%%%%%%%%%%%%%%%%%%%%%%%%%%%%%%%%%%%%%%%%
%%%%%%%%%%%%%%%%%%%%%%%%%%%%%%%%%%%%%%%%%%%%%%%%%%%%%%%%%%%%%%%%%%%%%%
%%%%%%%%%%%%%%%%%%%%%%%%%%%%%%%%%%%%%%%%%%%%
\section*{References} %%%%%%%%%%%%%%%%%%%%%%%%%%%%%%%%%%%%%%%%%%%%%%%%
%%%%%%%%%%%%%%%%%%%%%%%%%%%%%%%%%%%%%%%%%%%%%%%%%%%%%%%%%%%%%%%%%%%%%%

\end{document}